\documentclass[intlimits,twoside,a4paper]{article}
\usepackage{amsmath,amssymb}
\usepackage{graphicx}
\usepackage{wrapfig}
\usepackage[T2A]{fontenc}
\usepackage[cp1251]{inputenc}

\usepackage[eqsecnum]{cmpj2}

\issue{2012}{15}{1}{13701}

\doinumber{10.5488/CMP.15.13701}

\title[Ballistic spin filtering across the FM/SC interface]%
{Ballistic spin filtering across the ferromagnetic-semiconductor interface}
\author[Y.H. Li]{Y.H. Li\thanks{Email: liyh@semi.ac.cn }}
\authorcopyright{Y.H. Li, 2012}

\address{
State Key Laboratory for Superlattices and Microstructures, Institute of Semiconductors, \\Chinese Academy of Sciences, P.O. Box 912, Beijing 100083, China
}
\date{Received June 15, 2011, in final form November 30, 2011}
\begin{document}

\maketitle

\begin{abstract}
The ballistic spin-filter effect from a ferromagnetic metal into a semiconductor has theoretically been studied with an intention of detecting the spin polarizability of density of states in FM layer at a higher energy level. The physical model for the ballistic spin filtering across the interface between ferromagnetic metals and semiconductor superlattice is developed by exciting the spin polarized electrons into n-type AlAs/GaAs superlattice layer at a much higher energy level and then ballistically tunneling through the barrier into the ferromagnetic film. Since both the helicity-modulated and static photocurrent responses are experimentally measurable quantities, the physical quantity of interest, the relative asymmetry of spin-polarized tunneling conductance, could be extracted experimentally in a more straightforward way, as compared with previous models. The present physical model serves guidance for studying spin detection with advanced performance in the future.
\keywords spintronics, spin filtering, ballistic transport, tunneling conductance, semiconductor superlattice
\pacs 72.25.Dc, 73.40.-c
\end{abstract}

\section{Introduction}

Based on spin-polarized electrons as information carrier, spintronics has been developing as an attractive field for a new generation of electronic devices, in which the injection and detection of spin-polarized carriers has been recognized as a significant challenge~\cite{Pri98,Zut04}. The spin-dependent transport through ferromagnetic (FM)/semiconductor (SC) interfaces is an important technology in order to achieve room-temperature operation of both spin injection and spin detection~\cite{Mon98}.

The spin injection efficiency from FM to SC could be directly measured by using a built-in light emission diode (LED) structure and detecting its circular polarization degree of electric photoluminescence (e-PL)~\cite{Rug03,Man02,LiC05}. However, in the case of electron spin detection, where electron spins flow from the SC into the FM side, there was no photon emission detected. Therefore, the spin-dependent electron transport has been investigated using optical spin orientation technology~\cite{Hir00,Hir99,Tan03}. The photo-excited spin-polarized electrons have different transmission probabilities, depending on their spin orientation with respect to that of the ferromagnetic metal layer~\cite{Hir02,Zut04}. The measured spin-dependence of tunneling current relates to many other facts except for the intrinsic effect, for example, the photon energy of the exciting light~\cite{Ste04,Liu05,Mal02,Hir02} and the characteristic temperature~\cite{Tan06,Ade05}. In addition, the asymmetry in the absorption for the $\sigma^{+}$ and $\sigma^{-}$ lights could also be induced by magnetic circular dichroism (MCD) inside the FM layer itself~\cite{Tan03,Ste05}.

\subsection{Current theory}

\begin{wrapfigure}{i}{0.5\textwidth}
\centerline{\includegraphics[width=0.5\textwidth]{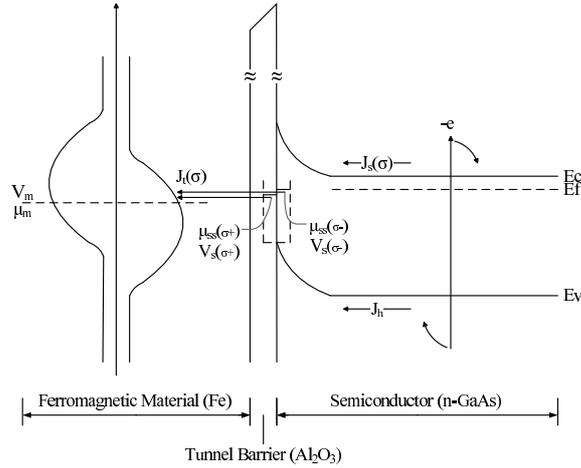}}
\caption{The band profile for a Fe/Al$_2$O$_3$/n-GaAs tunneling structure under forward bias. } \label{fig1}
\end{wrapfigure}

In the spin filtering process across FM/SC interface, the spin polarized ensemble excited by right or left circularly polarized light on the semiconductor side had to first relax down to the band edge, then transit through the tunneling barrier into the FM side~\cite{Jan98}. In the previous work, a physical model for the spin transport across Fe/Al$_2$O$_3$/n-GaAs tunneling structure under optical spin orientation was established~\cite{Xia08}. Figure~\ref{fig1} illustrates the band diagram for such a tunneling structure. The density of states for two spin opposite subbands are sketched on the metal side with a unique chemical potential ${\mu _{\mathrm{m}}}$. The state filling in the spin resolved interface states for two opposite spin orientations may have a dependence on the spin parity, leading to different chemical potentials, ${\mu _\mathrm{ss}}\left( {{\sigma ^ - }} \right)$ and ${\mu _\mathrm{ss}}\left( {{\sigma ^ + }} \right)$. Their potentials with respect to the Fermi level of the n-type semiconductor, which is taken as the potential zero, are labeled by ${V_\mathrm{s}}\left( {{\sigma ^ - }} \right)$ and ${V_\mathrm{s}}\left( {{\sigma ^ + }} \right)$, respectively.

The model system was based on a hypothesis that the Al$_2$O$_3$ barrier was not very transparent
so that its conductance $G_{\mathrm{t}}^\sigma$ was much smaller than the conductance $G_\mathrm{s}$ of Schottky barrier ($G_{\mathrm{t}}^\sigma  \ll {G_\mathrm{s}}$).
Either electrons or holes would be first captured by the interface states, and then escape through the tunneling barrier. The conclusion followed that the ratio of the helicity-modulated response to the chopped photocurrent was proportional to the sum of $\Delta {G_{\mathrm{t}}}/{G_{\mathrm{t}}}$, the relative asymmetry of spin-polarized tunneling conductance with $\Delta {G_{\mathrm{t}}} = \bar G_{\mathrm{t}}^ \uparrow  - \bar G_{\mathrm{t}}^ \downarrow $ being the conductance difference between spin-up and spin-down electrons, and the MCD effect from the ferromagnetic metal film.

There are some problems in such types of spin filtering mechanism. Firstly, it is quite conceivable that the energy relaxation makes the ensemble lose its initial spin polarization. Secondly, the injected spins on the metal side merely test the difference between spin-up and spin-down density of states (DOS) near the Fermi level, where the DOS difference between two spin orientations need not be the largest. All of these problems may lead to a diminutive spin filtering effect.

\subsection{Present principle and method}

\begin{figure}
\begin{minipage}[t]{0.5\linewidth}
\centering
\includegraphics[width=2.9in]{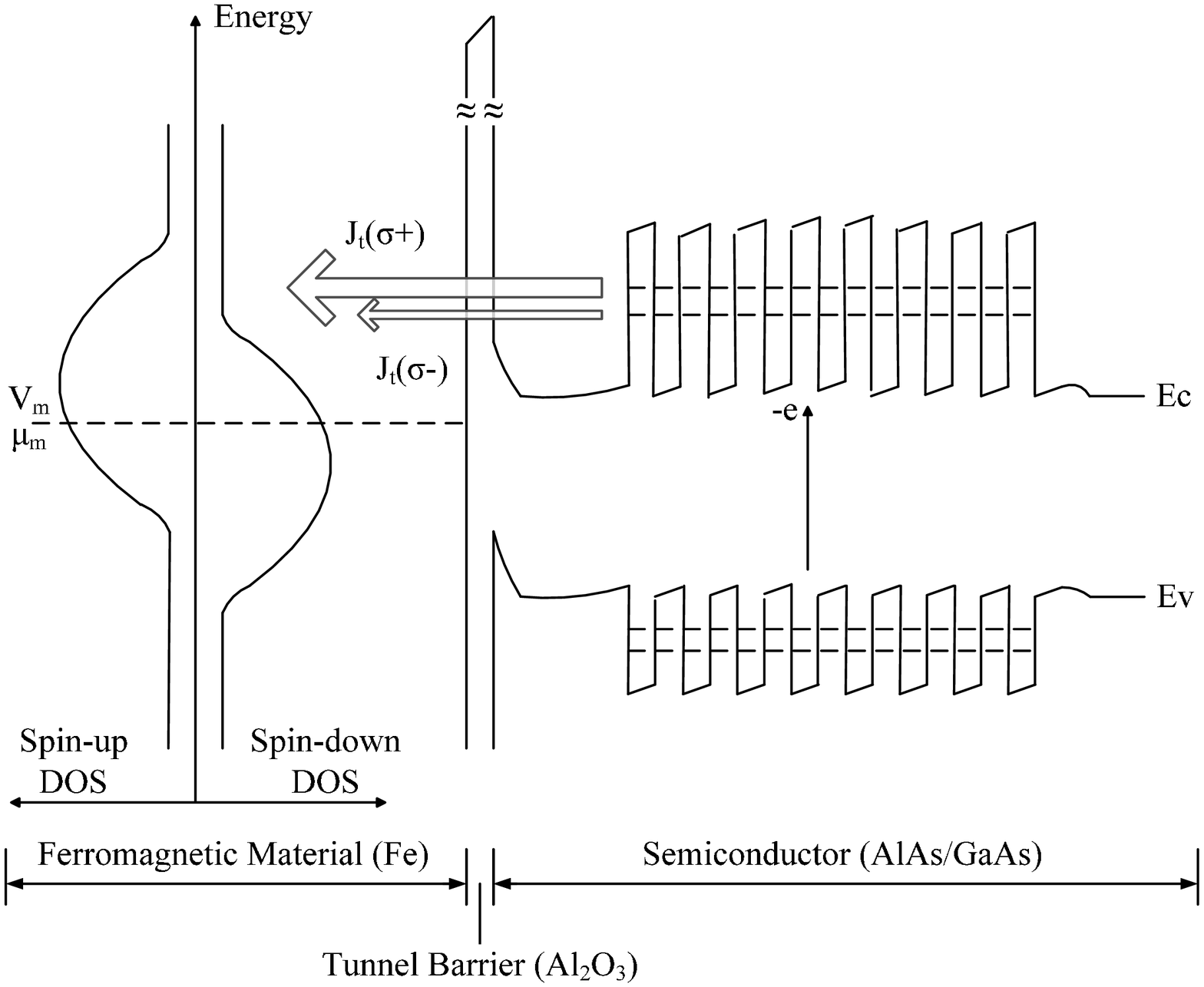}
\caption{Ballistic spin filtering across the interface between FM and a semiconductor superlattice structure at a higher energy level biased positively.}
\label{fig2}
\end{minipage}%
\begin{minipage}[t]{0.5\linewidth}
\centering
\includegraphics[width=2.9in]{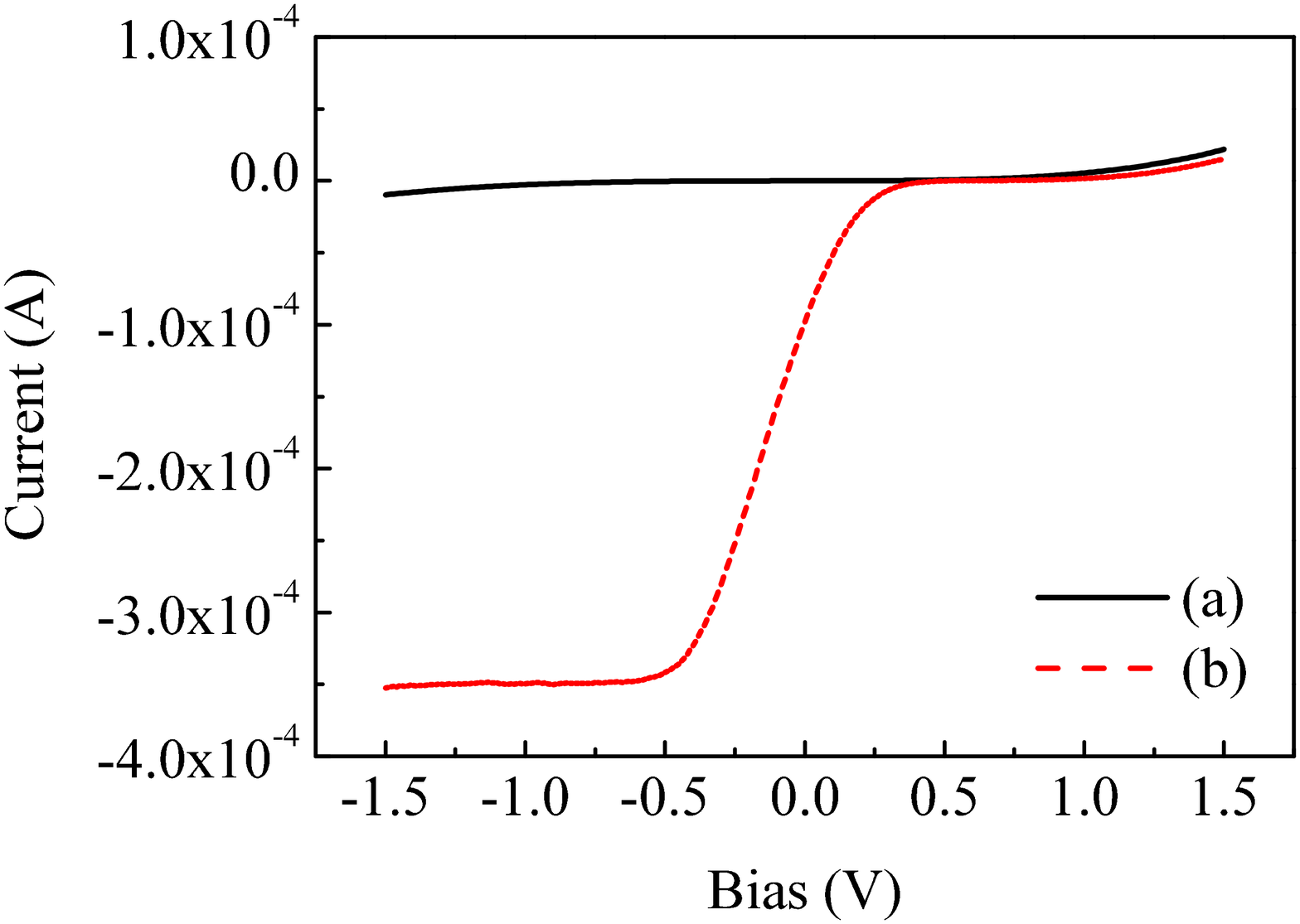}
\caption{The (a) dark and (b) light I-V curves of the sample structure including AlAs/GaAs superlattice.}
\label{fig3}
\end{minipage}
\end{figure}

To avoid the aforementioned drawbacks, a physical model was proposed for the ballistic spin filtering across the interface between FM and SC consisting in exciting the spin polarized electrons into n-type AlAs/GaAs superlattice layer at a much higher energy level and then ballistically tunneling through the barrier into FM side. The scheme is described in figure~\ref{fig2}, wherein the ${J_{\mathrm{t}}}\left( {{\sigma ^ - }} \right)$ and ${J_{\mathrm{t}}}\left( {{\sigma ^ + }} \right)$ represent the tunnel currents for two spin orientations, respectively, and the arrow broadness is proportional to the tunnel current denoting the probability of two transit processes. The resonant photo-excitation in the AlAs/GaAs superlattice creates spin-polarized electrons of high injection energy without significant energy relaxation. By positively biasing the structure, the spin polarized electrons are driven ballistically over a short distance ($<100$~\AA), and then hit the Al$_2$O$_3$ insulator barrier at an energy even higher than the thin Schottky barrier on the SC side. There is a profound need in studying the spin-filter effect in such novel structures. On the one hand, the present physical model intends to detect the spin-resolved DOS in FM layer at a higher energy level. On the other hand, the efficiency of spin filtering should be improved by ballistic transport.

Before going into detail, we must first be reminded that the I-V curves  would be quite different in the absence and in the presence of light. In the absence of illumination, the AlAs/GaAs superlattice behaves as an insulator, sharing a large portion of the applied voltage. By contrast, the AlAs/GaAs superlattice under the illumination becomes rather conductive so that the bias voltage mainly drops across the Al$_2$O$_3$ barrier layer. This is more obvious under a negative bias. The injected electrons from FM side transit over the AlAs/GaAs superlattice, leading to a substantial enhancement of the photocurrent at a negative bias. As a result, $\Delta {G_{\mathrm{t}}}/{\bar G_{\mathrm{t}}}$, the relative asymmetry of spin-polarized tunneling conductance is not the ratio of $\Delta {G_{\mathrm{t}}} = \bar G_{\mathrm{t}}^ \uparrow  - \bar G_{\mathrm{t}}^ \downarrow $ to the chopped photocurrent as it is in~\cite{Mal02,Xia08}, but to the  static photocurrent.

This point is very important to the theoretical derivation later on. To verify it experimentally, an AlAs(2~nm)/GaAs(1.8~nm) superlattice structure with a 10~nm undoped GaAs buffer layer was prepared on n$^+$-GaAs substrate by molecular beam epitaxy. On the top of them, the $3\div4$~ML thick aluminum film was deposited. Then the sample was taken out of the MBE vacuum chamber and placed in the atmosphere for a 24-hour natural oxidation in order to form the Al$_2$O$_3$ barrier layer. Afterwards, the sample was placed back into MBE chamber and covered with another 8~nm-thick iron film. The current-voltage characteristics of the sample with and without illumination are shown in figure~\ref{fig3}. A significant photocurrent response shows up under the negative bias. This observation is consistent with the  above statement. Therefore, it is the DC photo-conductance that should be taken as the denominator of $\Delta {G_{\mathrm{t}}}/{\bar G_{\mathrm{t}}}$.

The theoretic model of ballistic spin filtering through the structure of Fe/Al$_2$O$_3$/(GaAs/AlAs) superlattice under optical spin orientation is described in the following section.

\section{Theoretical model}

We describe the tunneling using the transfer Hamiltonian approach, which is a first-order perturbation method valid for the case of low tunnel barrier transparency. For the ferromagnetic electrode we define spin-dependent densities of states $\rho_{\mathrm{m}}^\sigma$ and an energy distribution function $f_{\mathrm{m}}$ independent of spin. The superscript $\sigma$ denotes the spin orientation with respect to a given quantization axis (either parallel $\uparrow$ or antiparallel $\downarrow$). For a (nonmagnetic) semiconductor superlattice, the density of states $\rho_{\mathrm{SL}}$ does not depend on spin. When no scattering centers are present in the tunnel barrier, the electron energy $\varepsilon$ and spin are conserved during the process of tunneling. The tunneling electron current ($J_{\mathrm{t}}^\sigma$) of spin orientation $\sigma$ from the semiconductor superlattice to the magnetic material is expressed as~\cite{Xia08}.
\begin{equation}
\label{itsigma}
J_{\mathrm{t}}^\sigma  = \frac{1}{e}\int {\rd\varepsilon \left[ {f_{\mathrm{SL}}^\sigma \left( \varepsilon  \right) - {f_{\mathrm{m}}}\left( {\varepsilon  + e{V_{\mathrm{m}}}} \right)} \right]} G_{\mathrm{t}}^\sigma \left( \varepsilon  \right).
\end{equation}
Here, ${f_{\mathrm{m}}}\left( {\varepsilon  + e{V_{\mathrm{m}}}} \right)$ and $f_{\mathrm{SL}}^\sigma \left( \varepsilon  \right)$ are the Fermi and quasi-Fermi distributions in the ferromagnetic metal and semiconductor superlattice, respectively. $V_{\mathrm{m}}$ is the bias voltage of the metal gate with respect to the quasi-Fermi level in the conduction band of superlattice. Formally, one can write down the following expression
\begin{equation}
\label{gtsigma}
G_{\mathrm{t}}^\sigma \left( \varepsilon  \right) = \frac{{2\pi {e^2}}}{\hbar }{\left| {{M^\sigma }\left( \varepsilon  \right)} \right|^2}{\rho _{\mathrm{SL}}}\left( \varepsilon  \right)\rho _{\mathrm{m}}^\sigma \left( {\varepsilon  + e{V_{\mathrm{m}}}} \right),
\end{equation}
 where $G_{\mathrm{t}}^\sigma(\varepsilon)$ depends on the densities of states $\rho_{\mathrm{SL}}(\varepsilon)$ in the semiconductor superlattice, the spin dependent densities of states $\rho _{\mathrm{m}}^\sigma \left( {\varepsilon  + e{V_{\mathrm{m}}}} \right)$ in the ferromagnetic metal, and $M^\sigma(\varepsilon)$, an energy- and spin-dependent tunneling matrix element that takes into account the overlapping of the wave functions of the respective materials.

In the present case, the filling in the DOS of FM film relevant to the tunneling is assumed to be zero, that is, ${f_{\mathrm{m}}}\left( {\varepsilon  + e{V_{\mathrm{m}}}} \right) \simeq 0$. As a result,
\begin{equation}
\label{itsigma2}
J_{\mathrm{t}}^\sigma  = \frac{1}{e}\int {\rd\varepsilon } f_{\mathrm{SL}}^\sigma (\varepsilon ) \cdot G_{\mathrm{t}}^\sigma (\varepsilon ).
\end{equation}
From rate equation inside the superlattice of a width of $W$ it follows that
\begin{align}
\label{itsigma3}
&I_{\mathrm{t}}^\sigma  = \frac{{e\eta P(\sigma )}}{{{E_{\mathrm{ph}}}}}\left( {1 - {\re^{ - \alpha W}}} \right) - \frac{{e\int {\rd\varepsilon {N_{\mathrm{SL}}}(\varepsilon )f_{\mathrm{SL}}^\sigma (\varepsilon )} }}{{{\tau _{\mathrm{R}}}}}\,,
\\
\label{intd}&
\int {\rd\varepsilon f_{\mathrm{SL}}^\sigma (\varepsilon )\left( {G_{\mathrm{t}}^\sigma (\varepsilon ) + \frac{{{e^2}{N_{\mathrm{SL}}}(\varepsilon )}}{{{\tau _{\mathrm{R}}}}}} \right)}  = \frac{{{e^2}\eta P(\sigma )}}{{{E_{\mathrm{ph}}}}}\left( {1 - {\re^{ - \alpha W}}} \right).
\end{align}
Since the energy range of DOS and their fillings in the superlattice are rather narrow, the integration on the left hand side can be replaced by $\left( {\bar G_{\mathrm{t}}^\sigma + \frac{{{e^2}{\bar N_{\mathrm{SL}}}}}{{{\tau _{\mathrm{R}}}}}} \right)\tilde \varepsilon _{\mathrm{R}}^\sigma$, where $\tilde \varepsilon _{\mathrm{R}}^\sigma$ denotes the quasi-Fermi level in the semiconductor superlattice for a specific spin orientation, ${\bar N_{\mathrm{SL}}}$ is the averaged constant DOS of the superlattice, $\bar G_{\mathrm{t}}^\sigma $ is the tunneling conductivity averaged over the energy range of the superlattice.
\begin{align}
\label{leftbar}
&\left( {\bar G_{\mathrm{t}}^\sigma  + \frac{{{e^2}{{\bar N}_{\mathrm{SL}}}}}{{{\tau _{\mathrm{R}}}}}} \right)\tilde \varepsilon _{\mathrm{R}}^\sigma  = \frac{{{e^2}\eta P(\sigma )}}{{{E_{\mathrm{ph}}}}}\left( {1 - {\re^{ - \alpha W}}} \right),
\\
\label{itsigma4}
&I_{\mathrm{t}}^\sigma  = \frac{1}{e}\int {\rd\varepsilon f_{\mathrm{SL}}^\sigma } (\varepsilon ) \cdot G_{\mathrm{t}}^\sigma (\varepsilon ) \approx \frac{{\bar G_{\mathrm{t}}^\sigma \tilde \varepsilon _{\mathrm{R}}^\sigma }}{e}\,,
\\
\label{itsigma5}
&I_{\mathrm{t}}^\sigma  \approx \dfrac{{\bar G_{\mathrm{t}}^\sigma \tilde \varepsilon _{\mathrm{R}}^\sigma }}{e} = \bar G_{\mathrm{t}}^\sigma \dfrac{{e\eta P(\sigma )\left( {1 - {\re^{ - \alpha W}}} \right)}}{{{E_{\mathrm{ph}}}\left( {\bar G_{\mathrm{t}}^\sigma  + \dfrac{{{e^2}{{\bar N}_{\mathrm{SL}}}}}{{{\tau _{\mathrm{R}}}}}} \right)}}\,.
\end{align}
When the tunneling barrier is not very transparent so that $\bar G_{\mathrm{t}}^\sigma  \ll \dfrac{{{e^2}{{\bar N}_{\mathrm{SL}}}}}{{{\tau _{\mathrm{R}}}}}$, one has
\begin{equation}
\label{itsigma6}
I_{\mathrm{t}}^\sigma  \approx \frac{{\bar G_{\mathrm{t}}^\sigma \tilde \varepsilon _{\mathrm{R}}^\sigma }}{e} \approx \bar G_{\mathrm{t}}^\sigma \frac{{e\eta P(\sigma )\left( {1 - {\re^{ - \alpha W}}} \right)}}{{{E_{\mathrm{ph}}}\left( {{e^2}{{\bar N}_{\mathrm{SL}}}/{\tau _{\mathrm{R}}}} \right)}}\,.
\end{equation}
It can be recognized that equation~(\ref{itsigma6}) relates the spin dependent tunneling current to $\bar G_{\mathrm{t}}^\sigma $, proportionally.

Next, one has to know the relevance of equation~(\ref{itsigma6}) to the experimentally measurable quantity. The helicity-related current measured using PEM and lock-in technology should be the difference between $I_{\mathrm{t}}^ \uparrow $ and $I_{\mathrm{t}}^ \downarrow $. To do that, one still has to consider the spin dependence of the light power, caused by MCD effect in FM film.

In MCD measurement, the mono-color light first passes a Rochon prism linear polarizer and becomes linearly polarized light with intensity of ${P_0}$. It can be described as the sum of two circularly polarized components ${P_0^{\sigma^+ }}$ and ${P_0^{\sigma^- }}$ that are in phase with each other:
\begin{equation}
\label{p0frac}
{P_0} = \frac{1}{2}\left( {P_0^{\sigma^+ } + P_0^{\sigma^- }} \right).
\end{equation}
Then, this linearly polarized light passes through a photoelastic modulator (PEM), which acts as a transparent, dynamically alternative quarter-wave plant at a modulation frequency of 50~kHz. The PEM will retard one of the components in equation~(\ref{p0frac}) (or advance the other component) with a time-dependent retardation, $\delta $, proportional to the modulator driving voltage.
\begin{equation}
\label{delta}
\delta  = {\delta _0}\sin \omega t,
\end{equation}
where ${\delta _0}$ is the peak retardation proportional to the modulator driving voltage of PEM, and is specified by each of PEM.

As is well-known, the MCD effect describes the phenomenon that the absorption of the matter to two opposite circularly polarized lights, the right-handed $\sigma ^ + $ and left-handed $\sigma ^ - $, is different due to the presence of either an external magnetic field or magnetization in the matter. The PEM-modulated light intensity after passing through the Fe metal film is given by
\begin{eqnarray}
P &=& \frac{{{P_0}}}{2}\left\{ {\left[ {1 - \sin ({\delta _0}\sin \omega t)} \right]{\re^{ - A({\sigma ^ - })}} + \left[ {1 + \sin ({\delta _0}\sin \omega t)} \right]{\re^{ - A({\sigma ^ + })}}} \right\}\nonumber\\
 &=& \frac{{{P_0}}}{2}\left\{ {\left[ {{\re^{ - A({\sigma ^ - })}} + {\re^{ - A({\sigma ^ + })}}} \right] + \left[ {{\re^{ - A({\sigma ^ + })}} - {\re^{ - A({\sigma ^ - })}}} \right]\sin ({\delta _0}\sin \omega t)} \right\}\nonumber\\
 &=& \frac{{{P_0}}}{2}\left[ {{\re^{ - A({\sigma ^ - })}} + {\re^{ - A({\sigma ^ + })}}} \right]\left[ {1 + \tanh \left( {\frac{{\Delta A}}{2}} \right)\sin \left( {{\delta _0}\sin \omega t} \right)} \right].
\label{pfrac}
\end{eqnarray}
The first term on the right hand side is the decayed intensity for the transmitted ${\sigma ^ - }$ polarized light, and the second one is for the transmitted ${\sigma ^ + }$ polarized light. $\Delta A = A({\sigma ^ - }) - A({\sigma ^ + })$ denotes the difference of the absorption rate between ${\sigma ^ - }$ and ${\sigma ^ + }$ polarized lights after they pass through FM film. When $\Delta A$ is small, say, less than 10\%,
\begin{eqnarray}
P &=& \frac{{{P_0}}}{2}\left[ {{\re^{ - A({\sigma ^ - })}} + {\re^{ - A({\sigma ^ + })}}} \right]\left[ {1 + \tanh \left( {\frac{{\Delta A}}{2}} \right)\sin \left( {{\delta _0}\sin \omega t} \right)} \right]\nonumber\\
 &=& {P_\mathrm{dc}}\left[ {1 + \tanh \left( {\frac{{\Delta A}}{2}} \right)\sin \left( {{\delta _0}\sin \omega t} \right)} \right]\nonumber\\
 &\simeq& {P_\mathrm{dc}}\left[ {1 + \frac{1}{2}\Delta A{\delta _0}\sin \omega t} \right],
\label{pfrac2}
\end{eqnarray}
where ${P_\mathrm{dc}}=\frac{1}{2}{P_0}\left( {{\re^{ - A({\sigma ^ - })}} + {\re^{ - A({\sigma ^ + })}}} \right)$ is static component of the transmitted light,
${P^\uparrow} =P_\mathrm{dc}\left( {1 + \frac{1}{2}\Delta A{\delta _0}} \right)$ and ${P^ \downarrow } = {P_\mathrm{dc}}\left( {1 - \frac{1}{2}\Delta A{\delta _0}} \right)$ at the peak modulation.
\begin{equation}
\label{ituparrow}
I_{\mathrm{t}}^ \uparrow  - I_{\mathrm{t}}^ \downarrow  \approx \bar G_{\mathrm{t}}^ \uparrow \frac{{e\eta P^ \uparrow \left( {1 - {\re^{ - \alpha W}}} \right)}}{{{E_{\mathrm{ph}}}\left( {{e^2}{{\bar N}_{\mathrm{SL}}}/{\tau _{\mathrm{R}}}} \right)}} - \bar G_{\mathrm{t}}^ \downarrow \frac{{e\eta P^ \downarrow \left( {1 - {\re^{ - \alpha W}}} \right)}}{{{E_{\mathrm{ph}}}\left( {{e^2}{{\bar N}_{\mathrm{SL}}}/{\tau _{\mathrm{R}}}} \right)}}\,.
\end{equation}
The AC signal detected by lock-in amplifier is equal to
\begin{eqnarray}
\tilde I_{\mathrm{t}}^{\mathrm{heli}} &=& I_{\mathrm{t}}^ \uparrow  - I_{\mathrm{t}}^ \downarrow \nonumber\\
 &\approx& \left( {\bar G_{\mathrm{t}}^ \uparrow {P^ \uparrow } - \bar G_{\mathrm{t}}^ \downarrow {P^ \downarrow }} \right)\frac{{e\eta \left( {1 - {\re^{ - \alpha W}}} \right)}}{{{E_{\mathrm{ph}}}\left( {{e^2}{{\bar N}_{\mathrm{SL}}}/{\tau _{\mathrm{R}}}} \right)}}\nonumber\\
 &=& \left[ {\left( {\bar G_{\mathrm{t}}^ \uparrow  - \bar G_{\mathrm{t}}^ \downarrow } \right) + \left( {\frac{{\bar G_{\mathrm{t}}^ \uparrow  + \bar G_{\mathrm{t}}^ \downarrow }}{2}} \right)\Delta A{\delta _0}} \right]\frac{{{P_\mathrm{dc}}e\eta \left( {1 - {\re^{ - \alpha W}}} \right)}}{{{E_{\mathrm{ph}}}\left( {{e^2}{{\bar N}_{\mathrm{SL}}}/{\tau _{\mathrm{R}}}} \right)}}\,.
\label{tildeit}
\end{eqnarray}
The second term stems purely from the MCD effect as a result of the different absorption of FM film to the right and left circularly polarized lights. Only the first term can be used to examine the spin dependent tunneling conductivities. Let ${\bar G_{\mathrm{t}}} = \frac{1}{2}\left( {\bar G_{\mathrm{t}}^ \uparrow  + \bar G_{\mathrm{t}}^ \downarrow } \right)$ be averaged tunneling conductivity.
\begin{equation}
\label{tildeit2}
\tilde I_{\mathrm{t}}^{\mathrm{heli}} = \left( {\frac{{\Delta {G_{\mathrm{t}}}}}{{{{\bar G}_{\mathrm{t}}}}} + \Delta A{\delta _0}} \right) \cdot {\bar G_{\mathrm{t}}}\frac{{{P_\mathrm{dc}}e\eta \left( {1 - {\re^{ - \alpha W}}} \right)}}{{{E_{\mathrm{ph}}}\left( {{e^2}{{\bar N}_{\mathrm{SL}}}/{\tau _{\mathrm{R}}}} \right)}}\,.
\end{equation}
Now, the question is how to extract the information on the spin dependent tunneling from experiments. It becomes clear later that
\begin{equation}
\label{itpbar}
I_{\mathrm{t}}^P = {\bar G_{\mathrm{t}}}\frac{{{P_\mathrm{dc}}e\eta \left( {1 - {\re^{ - \alpha W}}} \right)}}{{{E_{\mathrm{ph}}}\left( {{e^2}{{\bar N}_{\mathrm{SL}}}/{\tau _{\mathrm{R}}}} \right)}}
\end{equation}
should be the spin independent photocurrent.

Let us check the spin independent photocurrent excited by linearly polarized light, which is the sum of the photocurrents induced by right and left circularly polarized lights
\begin{eqnarray}
\sum\limits_\sigma  {I_{\mathrm{t}}^\sigma}  &\approx &\sum\limits_\sigma  {\bar G_{\mathrm{t}}^\sigma } \frac{{ e\eta P(\sigma )\left( {1 - {\re^{ - \alpha W}}} \right)}}{{{E_{\mathrm{ph}}}\left( {{e^2}{{\bar N}_{\mathrm{SL}}}/{\tau _{\mathrm{R}}}} \right)}}\nonumber\\
 &=& \frac{{e\eta \left( {1 - {\re^{ - \alpha W}}} \right)}}{{{E_{\mathrm{ph}}}\left( {{e^2}{{\bar N}_{\mathrm{SL}}}/{\tau _{\mathrm{R}}}} \right)}}\left( {\bar G_{\mathrm{t}}^ \uparrow {P^ \uparrow } + \bar G_{\mathrm{t}}^ \downarrow {P^ \downarrow }} \right).
\label{sumlimitssigma}
\end{eqnarray}
The intensities of the lights after transmitting the FM film are given as
\begin{equation}
\label{pleftuparrowright}
{P^ \uparrow } = \frac{{{P_0}}}{2}{\re^{ - {A^ \uparrow }}}, \qquad
{P^ \downarrow } = \frac{{{P_0}}}{2}{\re^{ - {A^ \downarrow }}}.
\end{equation}
Therefore
\begin{eqnarray}
I_{\mathrm{t}}^P &=& \sum\limits_\sigma  {I_{\mathrm{t}}^\sigma}\nonumber\\
 &=& \frac{{e\eta {P_0}\left( {1 - {\re^{ - \alpha W}}} \right)}}{{2{E_{\mathrm{ph}}}\left( {{e^2}{{\bar N}_{\mathrm{SL}}}/{\tau _{\mathrm{R}}}} \right)}}\left( {\bar G_{\mathrm{t}}^ \uparrow {\re^{ - {A^ \uparrow }}} + \bar G_{\mathrm{t}}^ \downarrow {\re^{ - {A^ \downarrow }}}} \right)\nonumber\\
 &\approx& \frac{{e\eta {P_0}{\re^{ - \bar A}}\left( {1 - {\re^{ - \alpha W}}} \right)}}{{{E_{\mathrm{ph}}}\left( {{e^2}{{\bar N}_{\mathrm{SL}}}/{\tau _{\mathrm{R}}}} \right)}} \cdot {{\bar G}_{\mathrm{t}}}\,.
\label{itpsumlimitssigma}
\end{eqnarray}
Here $\bar A \simeq \frac{1}{2}\left( {{A^ \uparrow } + {A^ \downarrow }} \right)$ and ${P_\mathrm{dc}} = {P_0}{\re^{ - \bar A}}$. Therefore, $I_{\mathrm{t}}^P$ in equation~(\ref{itpbar}) may be approximated by the static photocurrent response in a device of the same layer structure, when it is excited by a linearly polarized light. Eventually
\begin{equation}
\label{fractilde}
\frac{{\tilde I_{\mathrm{t}}^{\mathrm{heli}}}}{{I_{\mathrm{t}}^P}} = \frac{{\Delta {G_{\mathrm{t}}}}}{{{{\bar G}_{\mathrm{t}}}}} + \Delta A{\delta _0}\,.
\end{equation}
Again, $\Delta A{\delta _0}$ is determined by $\Delta A{\delta _0} = \left[ {\left( {1 + \frac{1}{2}\Delta A{\delta _0}} \right) - \left( {1 - \frac{1}{2}\Delta A{\delta _0}} \right)} \right] = \left( {{P^ \uparrow } - {P^ \downarrow }} \right)/{P_0}$.

Since the static photocurrents $I_{\mathrm{t}}^P$, the helicity-dependent photo-currents $\tilde I_{\mathrm{t}}^{\mathrm{heli}}$ and the MCD effect $\Delta A{\delta _0}$ are measurable physical quantities~\cite{Hir00,Hir99,Tan03,Ste05,Xia08}, the asymmetry of spin-polarized tunneling conductance $\Delta {G_{\mathrm{t}}}/{\bar G_{\mathrm{t}}}$ can be extracted from the measurements in the FM/SC tunneling structure. Compared to the expression for normal spin filtering $\tilde J_{\mathrm{t}}^{\mathrm{heli}}/J_{\mathrm{t}}^P = {G_{\mathrm{t}}}/2{G_\mathrm{s}} \cdot \left[ {\Delta {G_{\mathrm{t}}}/{{\bar G}_{\mathrm{t}}} + \Delta A{\delta _0}} \right]$~\cite{Xia08}, the present result, $\tilde I_{\mathrm{t}}^{\mathrm{heli}}/I_{\mathrm{t}}^P = \left[ {\Delta {G_{\mathrm{t}}}/{{\bar G}_{\mathrm{t}}} + \Delta A{\delta _0}} \right]$, is related to the quantity $\Delta {G_{\mathrm{t}}}/{\bar G_{\mathrm{t}}}$ in a more straightforward way.

\section{Conclusions}

In summary, the theoretical model of the ballistic spin filtering across the interface between ferromagnetic metal and semiconductor superlattice was developed by exciting the spin polarized electrons into n-type AlAs/GaAs superlattice layer at a much higher energy level and then ballistically tunneling through the barrier into the ferromagnetic film. It turns out that the ratio of the helicity-modulated photo-response to the static photo-response is equal to the sum of the relative asymmetry in the tunneling conductance between two opposite spin-polarized tunneling channels and the MCD effect from the ferromagnetic metal layer. Since both the helicity-modulated and static photocurrent responses are experimentally measurable quantities from the spin transport measurement, the physical quantity of interest $\Delta {G_{\mathrm{t}}}/{\bar G_{\mathrm{t}}}$ could be extracted quantitatively in the framework of the ballistic tunneling model. Compared with traditional models of spin-filter effect, the relation of the measurable quantities to $\Delta {G_{\mathrm{t}}}/{\bar G_{\mathrm{t}}}$ is more straightforward in the present ballistic tunneling model.

\section*{Acknowledgements}

The author expresses gratitude to Professor H.Z.~Zheng for the valuable guidance and inspiring discussion. This work was partly supported by the National Basic Research Program, the Special Research Programs of Chinese Academy of Sciences and the National Natural Science Foundation of China.

\ukrainianpart

\title{Балістичне спінове фільтрування через міжфазову границю
феромагнетик-напівпровідник}
\author{Й.Х. Лi}

\address{Центральна державна лабораторія з дослідження надграток і мікроструктур, Інститут напівпровідників,  Академія наук Китаю, а/с 912, 100083 Пекін, КНР
}

\makeukrtitle

\begin{abstract}
\tolerance=3000%
Ефект балiстичного спiнового фiльтрування з феромагнiтного металу в
напiвпровiдник теоретично дослiджено з намiром виявлення спiнової
поляризованостi густини станiв у феромагнiтному шарi при вищому енергетичному
рiвнi. Розвинуто фiзичну модель для балiстичного спiнового фiльтрування через
мiжфазову границю мiж феромагнiтними металами i напiвпровдниковою
суперграткою, в основі якої є збудження спiнополяризаваних електронів у шарi
AlAs/GaAs супергратки n-типу при набагато вищому енергетичному рiвнi з
подальшим балiстичним тунелюванням через бар'єр у феромагнiтну плiвку.
Оскiльки обидва спiрально-модульований i статичний фотовiдгуки є
експериментально вимiрювальними величинами, фiзична величина, яка нас
цiкавить, вiдносна асиметрiя спiнополяризованої тунельної провiдностi, могла б
бути виокремлена експериментально в бiльш прямий спосiб порiвняно з
попереднiми моделями. Дана фiзична модель зорiєнтована на вивчення спiнового
детектування з високою продуктивністю в майбутньому.
\keywords магнетоелектроніка, спінове фільтрування, балістичне
перенесення, тунельна провідність, напівпровідникова супергратка
\end{abstract}

\end{document}